\documentclass[]{fairmeta}
\usepackage{graphicx}
\usepackage{amsmath}
\usepackage{enumitem}
\usepackage{listings}      
\usepackage{amsmath}      

\usepackage{amssymb}     
\usepackage{hyperref}     
\usepackage{caption}      
\usepackage{subcaption}   
\usepackage{xcolor}      
\usepackage{booktabs}
\usepackage{tabularx}
\usepackage{tcolorbox}
\usepackage{titlesec}
\usepackage{algorithm}
\usepackage{algpseudocode}
\tcbuselibrary{breakable} 
\usepackage{makecell}
\usepackage{listings}
\usepackage{xcolor}
\usepackage{balance}

\lstdefinestyle{jsonblock}{
    basicstyle=\ttfamily\footnotesize,
    breaklines=true,
    frame=single,
    backgroundcolor=\color{gray!10},
    columns=fullflexible
}

\setlist[itemize]{align=parleft,left=0pt..1em}

\titlespacing*{\section}{0pt}{0.8\baselineskip}{0.5\baselineskip}
\titlespacing*{\subsection}{0pt}{0.8\baselineskip}{0.1\baselineskip}

\usepackage{seqsplit}
\usepackage{microtype}

\newcommand{\sln}{HPCAgentTester}

\title{\sln{}: A Multi-Agent LLM Approach for Enhanced HPC Unit Test Generation
}
\author[1\dagger,3]{Rabimba Karanjai}
\author[2]{Lei Xu}
\author[1]{Weidong Shi}

\affiliation[1]{University Of Houston}
\affiliation[2]{{Kent State University}}
\contribution[\dagger]{rkaranjai@uh.edu}

\abstract{
Unit testing in High-Performance Computing (HPC) is critical but challenged by parallelism, complex algorithms, and diverse hardware. Traditional methods often fail to address non-deterministic behavior and synchronization issues in HPC applications.  This paper introduces \sln{}, a novel multi-agent Large Language Model (LLM) framework designed to automate and enhance unit test generation for HPC software utilizing OpenMP and MPI. \sln{} employs a unique collaborative workflow where specialized LLM agents (Recipe Agent and Test Agent) iteratively generate and refine test cases through a critique loop. This architecture enables the generation of context-aware unit tests that specifically target parallel execution constructs, complex communication patterns, and hierarchical parallelism. We demonstrate \sln{}'s ability to produce compilable and functionally correct tests for OpenMP and MPI primitives, effectively identifying subtle bugs that are often missed by conventional techniques. Our evaluation shows that \sln{} significantly improves test compilation rates and correctness compared to standalone LLMs, offering a more robust and scalable solution for ensuring the reliability of parallel software systems.
}

\date{\today}

\begin{document}
\maketitle

\section{Introduction}

Unit testing is a cornerstone of modern software engineering, acting as a crucial safeguard by identifying and preventing bugs early in the development lifecycle~\cite{BuildingAIAgentstoA,nichols2024can,karanjai2024harnessing,nichols2024performance}. A well-maintained test suite serves as a living documentation, illustrating the intended code behavior and preventing regressions during software evolution. However, High-Performance Computing (HPC) presents unique challenges to traditional software development. HPC software often involves intricate algorithms and sophisticated parallel processing techniques to tackle computationally intensive tasks. The development, optimization, and maintenance of such parallel software, which frequently utilizes diverse hardware architectures and programming models, is exceptionally complex~\cite{nichols2024hpc}.

Large Language Models (LLMs) have recently emerged as a groundbreaking technology, demonstrating remarkable capabilities in comprehending and generating both human language and computer code. This potential is increasingly recognized in software engineering tasks, including the automated generation of unit tests, which can alleviate the burden of manual creation. While advancements have shown LLMs' utility in generating parallel code~\cite{nichols2024can, nichols2024hpc} and basic unit tests for HPC leveraging OpenMP and MPI~\cite{karanjai2024harnessing}, the systematic generation of comprehensive tests that cover complex parallel patterns, intricate data dependencies, and effectively utilize domain-specific feedback for refinement remains an active area of research. Specifically, ensuring that LLM-generated tests for HPC are not only syntactically correct but also semantically robust in verifying parallel constructs requires a more structured approach.

Unit testing high-performance parallel programs is considerably more complex than testing traditional sequential codes, primarily because of concurrency. Parallel programs, whether using OpenMP for shared-memory parallelism or MPI for distributed-memory message passing, inherently require the management of shared states and coordination of interactions between multiple threads or processes~\cite{jin2011high}. Issues such as data races, where multiple threads access and modify shared memory concurrently without proper synchronization, can lead to nondeterministic program behavior, making bugs notoriously difficult to identify and reproduce.

To address these challenges, we propose HPCAgentTester, a novel framework centered on a multi-agent LLM workflow for generating unit tests for HPC applications using OpenMP and MPI. This collaborative and iterative process aims to produce more robust and context-aware unit tests. 

Specifically, this research investigates the capabilities of our proposed framework through the following research questions:
\vspace{-0.12cm}
\begin{itemize}[leftmargin=*]
    \item \textbf{RQ1:} \textit{To what extent can \sln{}'s multi-agent workflow, incorporating an iterative critique loop, effectively generate compilable and functionally correct unit tests for C++ HPC codes using OpenMP and MPI?}
    \item \textbf{RQ2:} \textit{How comprehensively does \sln{} capture and generate tests for a diverse range of OpenMP and MPI primitives compared to less structured LLM prompting techniques?}
    \item \textbf{RQ3:} \textit{How effectively does \sln{} generate unit tests that target and validate hierarchical parallelism, such as nested OpenMP regions and complex MPI communication patterns?}
\end{itemize}

\section{Related Work and New Contributions}\label{sec-related}

The use of Large Language Models (LLMs) for software development tasks, including code generation and automated testing, has grown rapidly in recent years. Several studies have explored their application in unit test generation, particularly in the context of high-performance computing (HPC).

Karanjai et al.~\cite{karanjai2024harnessing} evaluated LLMs such as Davinci and ChatGPT for generating unit tests for C++ code utilizing OpenMP and MPI. While the models produced mostly correct tests, common issues included redundant assertions and incomplete test logic~\cite{karanjai2024harnessing}. \sln{} builds upon these insights by introducing a structured, multi-agent workflow to overcome such limitations via explicit test criteria generation and an iterative critique mechanism.

Nichols et al. introduced \textit{ParEval}~\cite{nichols2024performance}, a benchmarking suite to assess LLM-generated parallel code. Their results highlight the difficulty LLMs face in producing correct and efficient MPI code, often due to a lack of runtime context and deep understanding of parallel semantics. Unlike ParEval, \sln{} targets unit test generation rather than code synthesis, focusing on verifying the correctness of pre-existing HPC programs, whether written by humans or generated by LLMs. Our emphasis on testing specific OpenMP and MPI constructs addresses the semantic gaps identified in ParEval.

Matija et al.~\cite{matija2024openmp} evaluated LLMs for OpenMP parallelization support, showing that models like ChatGPT and GitHub Copilot can assist developers in introducing parallel constructs. Although their work focused on transformation rather than testing, their analysis of LLM understanding of OpenMP is relevant to \sln{}’s design, which leverages similar knowledge to define testable behaviors.

Munley et al.~\cite{munley2023openacc} investigated LLMs for generating validation suites for HPC compilers, particularly OpenACC. Their focus was compiler-level verification, contrasting with \sln{}’s emphasis on functional validation of HPC application code. This distinction entails different goals, such as test granularity and domain-specific oracles.

A notable advance in LLM-assisted HPC is \textit{HPC-Coder}~\cite{nichols2024can}, a fine-tuned model specialized for HPC and scientific computing. HPC-Coder excels in tasks such as annotating OpenMP pragmas and generating context-aware code. While \sln{} is model-agnostic, it could benefit from integration with domain-specialized models like HPC-Coder. Our contribution lies not in building a new foundation model, but in orchestrating a modular pipeline of LLM agents, responsible for analysis, test specification, generation, and critique—to create semantically rich unit tests.

While prior work demonstrates LLM potential across various HPC tasks, from code generation and parallelization to compiler testing, \sln{} introduces a framework that directly addresses the complexity of unit testing in parallel programs. 

\underline{Key contributions of this work include:}
\begin{itemize}[nosep,leftmargin=*]
    \item A modular {\bf multi-agent architecture} that separates code analysis, test recipe generation, test code synthesis, and iterative critique;
    \item A refinement loop that improves semantic accuracy and coverage of tests through LLM-based feedback; and
    \item A focus on testing intricate OpenMP and MPI constructs, guided by formal test specifications rather than ad-hoc prompts.
\end{itemize}

Our framework complements existing LLM capabilities by applying structured reasoning and role-based agent coordination to the specific problem of HPC unit test generation.

\section{The Challenge of Unit Testing Parallel Code}
\label{sec:testing_challenge}

Unit testing traditional sequential programs is a well-understood discipline. A test typically provides a known input to a function and asserts that the output matches a single, predictable result. However, this paradigm is insufficient for High-Performance Computing (HPC) applications that rely on parallel programming models like the Message Passing Interface (MPI) for distributed-memory systems and OpenMP for shared-memory systems. The introduction of concurrency brings forth new classes of bugs that do not exist in sequential code and which are notoriously difficult to reproduce and debug \cite{abbaspour2017concurrency,hochstein2011isi}. Effective unit testing in HPC must therefore evolve to validate not just the functional correctness of an algorithm, but also the correctness of its parallel execution.

\subsection{Illustrating a Testing Challenge in MPI}
MPI enables parallelism by allowing independent processes, each with its own memory, to communicate by sending and receiving messages. A fundamental challenge is ensuring that these communication patterns are correct and free from deadlocks.
Consider a simple scenario where two processes need to exchange data. A naive implementation might look like this:

\begin{figure}[t]
\begin{lstlisting}[
  language=C++,
  basicstyle=\scriptsize\ttfamily,
  breaklines=true,
  frame=single,
  framerule=0.4pt,
  numbers=left,
  caption={Simplified C++ MPI code for symmetric send/receive, which may deadlock.},
  label=lst:mpi_deadlock
]
// Simplified C++ with MPI
void exchange_data(int rank, int partner_rank) {
    int send_buf = rank;
    int recv_buf = -1;
    // Both processes try to send first
    MPI_Send(&send_buf, 1, MPI_INT, partner_rank, 0, MPI_COMM_WORLD);
    MPI_Recv(&recv_buf, 1, MPI_INT, partner_rank, 0, MPI_COMM_WORLD, MPI_STATUS_IGNORE);
    // ...
}
\end{lstlisting}
\end{figure}
This code contains a classic deadlock bug. Since `MPI\_Send` is a blocking call, both processes will wait indefinitely for the other to call `MPI\_Recv` first, which will never happen. A correct implementation requires an ordered send/receive pattern (e.g., rank 0 sends then receives, while rank 1 receives then sends).

A unit test for this scenario must therefore not only check if the data was exchanged correctly but also verify that the communication completes at all. A conceptual test using a framework like Google Test might be structured as:

\begin{figure}[t]
\begin{lstlisting}[
  basicstyle=\scriptsize\ttfamily,
  breaklines=true,
  frame=single,
  framerule=0.4pt,
  captionpos=b,
  caption={Example MPI test case for deadlock detection with timeout.},
  label=fig:mpi_test
]
TEST(MPI_Bugs, DeadlockOnSymmetricSend) {
    // Setup: Get MPI rank and size
    int rank, size;
    MPI_Comm_rank(MPI_COMM_WORLD, &rank);
    MPI_Comm_size(MPI_COMM_WORLD, &size);
    ASSERT_EQ(size, 2) << "This test requires exactly 2 processes.";

    // This function call is expected to hang, so a test
    // would need a timeout mechanism to detect the deadlock.
    // The test *passes* if it correctly identifies the hang.
    ASSERT_TIMEOUT(exchange_data(rank, 1-rank), 5.0);
}
\end{lstlisting}
\end{figure}

This simple example highlights the need for tests that understand and can verify communication protocols, synchronization points, and potential deadlocks—concerns that are central to \sln{}'s design.

\subsection{Testing Challenge in OpenMP}
OpenMP simplifies on-node parallelism by using compiler directives (`pragmas`) to distribute work across multiple threads that share memory. While this avoids explicit message passing, it introduces the risk of data races. A data race occurs when multiple threads access the same memory location concurrently, and at least one of those accesses is a write, without proper synchronization.

Consider a parallel loop to compute the sum of an array:
\begin{figure}[t]
\begin{lstlisting}[
  language=C++,
  basicstyle=\scriptsize\ttfamily,
  breaklines=true,
  frame=single,           % Optional: Adds a box around the code
  numbers=left,           % Optional: Adds line numbers
  escapechar=@            % Critical: Allows # in code via @#
  caption={Simplified C++ OpenMP code demonstrating a race condition in a parallel sum. The pragma must be escaped as @#pragma.},
  label=lst:omp_race_sum
]
// Simplified C++ with OpenMP
double parallel_sum(double* data, int size) {
    double total = 0.0;
    @#pragma omp parallel for
    for (int i = 0; i < size; ++i) {
        total += data[i]; // BUG: Race condition on 'total'
    }
    return total;
}
\end{lstlisting}
\end{figure}

Here, multiple threads read and write to the shared variable \texttt{total} simultaneously without protection. The final result is non-deterministic and almost certainly incorrect. The fix involves using a \texttt{reduction} clause (\texttt{\#pragma omp parallel for reduction(+:total)}), which instructs the compiler to generate safe code for this pattern.

A unit test for `parallel\_sum` must do more than check for a single correct value; it must validate that the parallel computation is correct and consistent.
\begin{figure}[t]
\begin{lstlisting}[
  language=C++,
  basicstyle=\scriptsize\ttfamily,
  breaklines=true,
  frame=single,           % (optional: adds a box)
  numbers=left,           % (optional: line numbers)
  escapechar=@,           % (critical for OpenMP pragmas)
  caption={C++ test case detecting a race condition in a parallel sum. 
           The test passes if `parallel\_sum` is consistent.},
  label=lst:omp_race_test
]
TEST(OpenMP_Bugs, DetectsRaceConditionOnSum) {
    // Setup: Create input data and calculate expected result sequentially
    std::vector<double> data = {1.0, 2.0, 3.0, 4.0};
    double expected_sum = 10.0;

    // Execute the parallel function multiple times
    bool is_consistent = true;
    for (int i = 0; i < 100; ++i) {
        if (parallel_sum(data.data(), data.size()) != expected_sum) {
            is_consistent = false;
            break;
        }
    }
    // The test passes if the parallel version consistently fails to produce
    // the correct result, indicating a likely race condition. A more
    // sophisticated test would use tools to detect the race directly.
    ASSERT_FALSE(is_consistent) << "Race condition not detected or function is correct.";
}
\end{lstlisting}
\end{figure}

This illustrates that OpenMP testing must focus on data scoping (e.g., `shared`, `private` variables), synchronization (`critical`, `atomic`), and work-sharing constructs to ensure parallel correctness.

\subsection{MPI Testing Considerations}

Message Passing Interface (MPI) programs rely heavily on correct communication patterns across independent processes. \sln{} targets three major aspects: First, point-to-point communication is tested by verifying that message transmission and reception via routines like \texttt{MPI\_Send} and \texttt{MPI\_Recv} succeed under varying conditions, such as blocking and non-blocking modes, data types, and rank configurations~\cite{compatmpi}. Second, collective operations,including \texttt{MPI\_Bcast}, \texttt{MPI\_Scatter}, \texttt{MPI\_Gather}, and \\\texttt{MPI\_Reduce},are tested to ensure correctness across different communicator sizes and in edge cases, such as degenerate one-process communicators~\cite{AutomatedMPI}. Third, environment management is validated through the correct usage of \texttt{MPI\_Init} and \texttt{MPI\_Finalize}, as well as retrieval of communicator metadata using  \texttt{MPI\_Comm\_rank} and \texttt{MPI\_Comm\_size}~\cite{gropp2003parallel}.

\subsection{OpenMP Testing Considerations}

OpenMP programs depend on shared-memory parallelism driven by compiler directives and runtime routines. \sln{} ensures correct behavior in five key domains. Parallel region execution is validated by confirming that multiple threads are spawned as expected through \texttt{\#pragma omp parallel}, with accurate detection of thread identities via \texttt{omp\_get\_thread\_num()} and team sizes using \texttt{omp\_get\_num\_threads()}. For work-sharing constructs, the distribution of loop iterations under directives such as \texttt{\#pragma omp for} is tested under various scheduling policies (e.g., static, dynamic, guided), along with full coverage of all sections in \texttt{\#pragma omp sections} blocks~\cite{UsingOpenMP}. Synchronization correctness is assessed by testing mutual exclusion through \texttt{\#pragma omp critical}, atomic operations via \texttt{\#pragma omp atomic}, and thread coordination with \texttt{\#pragma omp barrier}. Data scoping rules are evaluated by ensuring correct treatment of \texttt{shared}, \texttt{private}, and \texttt{firstprivate} variables, confirming proper isolation and initialization semantics~\cite{maclaren2011introduction}. Finally, task-based parallelism is tested through the correct creation and execution of tasks using \texttt{\#pragma omp task}, enforcement of dependency constraints via the \texttt{depend} clause, and synchronization using \texttt{taskwait}~\cite{wang2012openmp}.

This behavioral coverage ensures that unit tests generated by \sln{} not only validate syntactic correctness but also probe the semantic correctness of parallel constructs central to HPC program execution.

\section{\sln{} Architecture}\label{sec:architecture}

\begin{figure}
    \centering
    \includegraphics[width=0.8\linewidth,height=3.6cm]{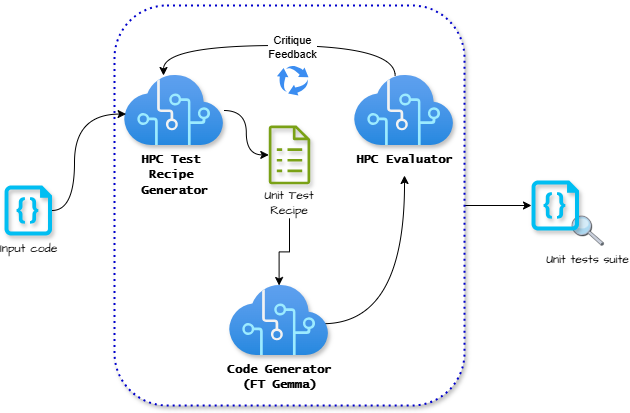}
    \caption{\sln{} system architecture. The workflow includes a code analyzer, recipe generation, test synthesis, and iterative critique, enabling robust HPC unit test generation.}
    \label{fig:overview}
\end{figure}

\sln{} is designed around a collaborative, multi-agent workflow that systematically translates high-level HPC source code into a verifiable suite of unit tests. Its architecture, visualized in Figure~\ref{fig:overview}, emphasizes the separation of concerns: code analysis, test strategy formulation (the "recipe"), test code generation, and iterative refinement. This structure allows each specialized agent to handle a distinct part of the complex test generation task, with a crucial feedback loop ensuring the final output is robust and correct. A key design tenet is that the framework generates unit tests for user-developed HPC application code, focusing on its functional and parallel correctness, rather than testing the underlying MPI or OpenMP runtime implementations themselves.


\begin{figure}[t]
    \centering
    \includegraphics[width=0.85\linewidth,height=4.8cm]{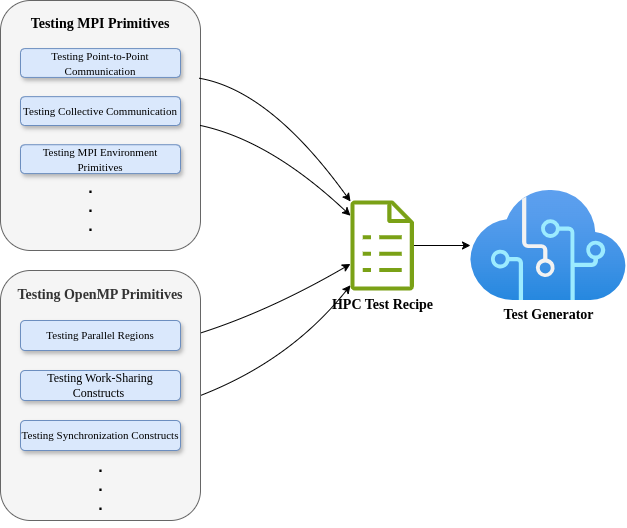}
    \caption{Test recipe generation by the Recipe Agent. Based on metadata from static analysis, the agent defines what constructs to test, the testing method, and the rationale behind each case.}
    \label{fig:test-rec}
\end{figure}

\subsection{Input HPC Code Analyzer (Code Analyzer)}
\label{sec:codeanalyze}

The first component of \sln{} is the Input HPC Code Analyzer. It performs a multi-faceted static analysis of the input HPC code (C++ with OpenMP/MPI) to extract features relevant to its parallel structure and potential concurrency vulnerabilities. The process, outlined in Algorithm 1, produces rich, structured metadata that forms the foundation for all subsequent stages.


The overall process, outlined in Algorithm~\ref{alg:code_analyzer}, produces rich metadata for subsequent stages.

\begin{algorithm}[t]
\caption{Input HPC Code Analysis Process}
\label{alg:code_analyzer}
\scriptsize
\begin{algorithmic}[1]
\Require HPC source code $S_{hpc}$
\Ensure Structured metadata $M_{out}$ (JSON format)
\State $AST \gets \textproc{Parse}(S_{hpc}, \text{Tree-sitter})$
\State $ClangTidyIssues \gets \textproc{AnalyzeWithClangTidy}(S_{hpc})$
\State $OpenMP \gets \textproc{ExtractOpenMPDirectives}(AST)$
\State $MPI \gets \textproc{ExtractMPICalls}(AST)$
\State $ParallelConstructs \gets OpenMP \cup MPI$
\State $DataFlow \gets \textproc{AnalyzeDataDependencies}(AST,$
\Statex $ParallelConstructs)$
\State $ControlFlow \gets \textproc{AnalyzeControlFlow}(AST)$
\State $Functions \gets \textproc{ExtractFunctionSignatures}(AST)$
\State $TestingAreas \gets \emptyset$
\State $KG \gets \textproc{LoadHPCBugKnowledgeGraph}()$
\ForAll{$c \in ParallelConstructs$}
    \ForAll{$p \in \textproc{QueryKnowledgeGraph}(KG, \text{type\_of}(c))$}
        \If{$c$ matches $p$ structurally or semantically}
            \State Add $(c, \textproc{description}(p), \textproc{test\_type}(p))$ to $TestingAreas$
        \EndIf
    \EndFor
\EndFor
\State $M_{out} \gets \textproc{FormatAsJSON}(\{$%
\Statex \hspace{1.7em} $AST, ParallelConstructs, DataFlow, ControlFlow,$
\Statex \hspace{1.7em} $TestingAreas, ClangTidyIssues, Functions \})$
\State \Return $M_{out}$
\end{algorithmic}
\end{algorithm}

The analyzer first uses standard tool \texttt{clang-tidy} \cite{clangtidy} for initial error checking. The core of the analysis, however, relies on custom parsers built with \texttt{tree-sitter} \cite{treesitter}. These parsers enable precise, language-aware extraction of parallel constructs and their context. 

Static analysis alone cannot definitively prove the existence of dynamic, run-time bugs like deadlocks, which often depend on input data and process scheduling. Instead, the analyzer's goal is to identify code patterns that are known to be susceptible to such issues (\texttt{MPI\_Barrier} inside a conditional block). These identified "areas of interest" are then passed to the Recipe Agent for targeted test generation.

\subsubsection{Identification of Potential Testing Areas using an HPC Bug Knowledge Graph}
A cornerstone of the Analyzer is its use of a novel \textbf{HPC Bug Knowledge Graph (KG)}. The development and curation of this KG represent a core contribution of our work. It is designed as a rich repository of common HPC bugs, anti-patterns, and testing heuristics. The KG is constructed via a semi-automated process combining automated data mining with expert validation:
\begin{itemize}[leftmargin=*]
    \item \textbf{Automated Information Extraction:} We use natural language processing (NLP) scripts to parse sources like the official MPI and OpenMP standards documents, public bug repositories of major HPC projects (PETSc, Trilinos), and academic papers on HPC software defects. This process extracts relationships between specific API calls, code structures, and documented bugs (e.g., `MPI\_Send` without a matching `MPI\_Recv` is linked to "potential deadlock").
    \item \textbf{Expert Curation and Structuring:} The automatically extracted patterns are then reviewed, structured, and validated by human experts in HPC. This manual step ensures the accuracy of the bug patterns, removes false positives, and enriches the KG with heuristics and testing strategies that are difficult to extract automatically.
\end{itemize}
When the Analyzer identifies a parallel construct, it queries the KG to find matching bug patterns. For instance, identifying a shared variable being modified in an OpenMP loop without a \texttt{critical} or \texttt{atomic} pragma would match a known "data race" pattern in the KG. The KG then returns a recommendation to generate a test specifically designed to trigger and detect this race condition.

When the Analyzer identifies a parallel construct or pattern, it queries the KG with the construct type and its surrounding context (data sharing attributes, enclosing loops). The KG might return patterns describing potential bugs (data races, deadlocks, incorrect collective operation usage) if specific conditions are met or safeguards are missing. For instance, identifying an \texttt{MPI\_Barrier} within a conditional block reached by only a subset of processes in a communicator would trigger a KG query; the KG might return a high-priority pattern for potential deadlock. This KG-driven approach allows for more nuanced, context-aware, and prioritized identification of testing areas than relying on a fixed list of hardcoded error patterns.

\subsubsection{Outputs}
The analyzer generates a structured representation of the analyzed code, typically in JSON format. This output meticulously details identified OpenMP and MPI constructs, data dependencies, function interfaces, control flow summaries, and, importantly, explicitly flags potential areas for focused testing, annotated with insights (bug type, severity) derived from the KG-informed pattern matching.

\subsection{Unit Test Recipe Generator (Recipe Agent)}
The Recipe Agent is an LLM-based component that acts as an expert test strategist. It translates the metadata from the Code Analyzer into a detailed, structured "Test Recipe"—an explicit, human-readable test plan in JSON format. 

This agent's effectiveness stems from its specialized fine-tuning. We created a curated dataset for this purpose, comprising triplets: `(code\_construct, bug\_pattern, test\_strategy)`. These triplets are derived from our HPC Bug KG, established HPC textbooks, and official programming guides. Fine-tuning an LLM (Gemma, Llama 2) on this dataset enhances its ability to reason about parallel correctness.

The agent's inputs are: 1) the JSON output from the Code Analyzer, and 2) its internal, fine-tuned knowledge of HPC testing. Its output, the Test Recipe, is a detailed blueprint outlining the exact functions to test, the specific parallel conditions to create ( number of threads, message sizes), the precise assertions to make, and a justification for each test case rooted in the analysis. This structured recipe serves as a critical grounding mechanism, constraining the next agent to generate meaningful and targeted code.


The Recipe Agent is designed to annotate key decisions within the recipe with brief justifications. For example:
\begin{lstlisting}[style=jsonblock,caption={Recipe example.},label={lst:mpi_deadlock_json},
language=]
{
  "test_id": "RECIPE_MPI_DEADLOCK_001",
  "target_construct": "MPI_Send_line_55_MPI_Recv_line_72",
  "test_type": "MPI_Potential_Deadlock_Order_Mismatch",
  "conditions": {
    "num_processes": 2,
    "rank0_send_first": true,
    "rank1_recv_first": true 
  },
  "expected_behavior_under_test": "Test may hang...",
  "justification_notes": [
    {"source": "Analyzer", "detail": "Identified MPI_Send..."},
    {"source": "KG_Pattern", "id": "KGP_MPI_015", ...},
    {"source": "Constraint_DB", "rule_id": "CDR_MPI_SYNC_003", ...}
  ],
  "suggested_assertion_method": "Verify completion..."
}
\end{lstlisting}
\vspace{0.3cm}

This explicit rationale makes the "why" behind each test clearer to developers, aids in debugging the test generation process itself, and allows for more informed review of the proposed test plan.

\subsection{Unit Test Case Generator Agent (Test Agent)}
The Test Agent is responsible for translating the abstract Test Recipe into executable C++ unit test code using a suitable testing framework (Google Test). This agent employs a powerful code-generation LLM (we used Gemma-2 in our experiments) that has been fine-tuned on the publicly available HPC-Instruct dataset. 
This domain-specific fine-tuning improves its fluency in generating code that correctly uses OpenMP and MPI APIs within a parallel context.

The Test Agent takes two inputs: 1) the structured Test Recipe (JSON) from the Recipe Agent, and 2) the original HPC source code. Its output is a complete, compilable C++ test file, including all necessary setup/teardown logic (`MPI\_Init`/`Finalize`), test case logic exercising the parallel behavior specified in the recipe, and assertion statements to verify the outcomes.


The Test Agent's LLM is fine-tuned using the HPC-Instruct dataset, which is specifically designed to enhance its ability to generate code that correctly utilizes OpenMP and MPI APIs, adheres to common HPC coding patterns, and understands parallel execution contexts. This domain-specific fine-tuning aims to improve the quality, correctness, and relevance of the generated tests beyond what a general-purpose code generation model might produce. 


\subsection{Feedback and Critique Loop}
\label{sec:feedback_loop}
A critical part of \sln{} is its iterative critique loop, designed to mitigate common LLM errors like hallucinations or omissions. Instead of being a final output, the code generated by the Test Agent is passed back to the Recipe Agent, which now assumes the role of a "critic."

Leveraging its understanding of the original test intent (as encoded in the Test Recipe), the critic agent analyzes the generated code for: \textbf{Adherence to Recipe:} Does the test implement all specified conditions and assertions?
\textbf{Correctness:} Is the test logic itself sound? Does it correctly use the testing framework and parallel APIs? and
\textbf{Relevance:} Does the test appear to genuinely exercise the intended part of the source code?

If flaws are found, the critic provides structured feedback ("The MPI\_Bcast test must include an assertion on every rank, not just rank 0.") to the Test Agent, which then attempts to regenerate the code. This cycle repeats until the test meets the recipe's criteria or a fixed number of iterations is reached, at which point it can be flagged for human review. This self-correction mechanism significantly improves the quality and reliability of the generated tests.

\subsubsection{Feedback, Iteration, Confidence Scoring, and Human Oversight}
The Recipe Agent (as critic) provides structured, actionable feedback to the Test Agent. This feedback can take several forms:
\begin{itemize}[noitemsep,topsep=0pt,parsep=0pt,partopsep=0pt] 
    \item \textbf{Natural Language Instructions:} Guiding the Test Agent with specific improvements ("The MPI\_Bcast test does not verify data on all ranks; ensure assertions are checked by all participating processes.").
    \item \textbf{A Modified/Refined Test Recipe:} Adjusting the original recipe with more explicit constraints or breaking down a complex test into simpler, more manageable parts if the Test Agent struggles.
    \item \textbf{Standardized Error Codes or Flags:} Representing specific, common issues \texttt{ERR\_RECIPE\_CONSTRAINT\_VIOLATED:NUM\_THREADS}, \texttt{WARN\_ASSERTION\_TARGET\_MISMATCH}, \\\texttt{SUGGEST\_RETRY\_WITH\_NON\_BLOCKING\_MPI}).
\end{itemize}
The Test Agent uses this feedback to regenerate or modify the test code. This iterative process continues until the Recipe Agent is satisfied with the quality and coverage of the generated tests or until a predefined number of attempts (five, determined empirically to prevent infinite loops) has been made.

To make this process more nuanced and manage inherent LLM uncertainties, the Recipe Agent's feedback includes a \textbf{confidence score} for its critique. This score, potentially derived from factors like the directness of a Test Recipe constraint violation (high confidence if a required setup like \texttt{MPI\_Init} is missing), the ambiguity of the generated test's logic (lower confidence for subtle semantic flaws), or the historical success rate of the Test Agent in implementing similar recipe patterns, indicates the critic's certainty about the identified issue. Persistently low-confidence critiques or tests that fail to be corrected after several iterations might be flagged with their history of critique and automatically escalated for \textbf{human review}, incorporating an active learning or human-in-the-loop element. This approach explicitly acknowledges that fully automated generation for all complex HPC scenarios is challenging and provides a mechanism for expert intervention when needed, thereby enhancing trustworthiness.

\subsubsection{Addressing LLM Biases and Errors}
The multi-agent design, centered around the structured Test Recipe and the iterative critique loop, is \sln{}'s primary strategy for addressing potential LLM biases and errors:
\begin{itemize}[noitemsep,topsep=0pt,parsep=0pt,partopsep=0pt] 
    \item \textbf{Specialization and Focused Expertise of Agents:} The Recipe Agent is fine-tuned for HPC testing strategy and understanding parallel semantics, while the Test Agent is fine-tuned for HPC code generation. This division of labor allows each agent to develop more focused expertise.
    \item \textbf{Structured Intermediate Representation as a Grounding Mechanism:} The Test Recipe—especially an explainable one—acts as a critical, human-understandable intermediary, constraining the Test Agent.
    \item \textbf{Iterative Refinement with Justification and Confidence:} The critique loop, with feedback, explanations, and confidence scores, guides the Test Agent towards better versions.
    \item \textbf{Diverse Knowledge Sources:} Relying on multiple sources (standards, KG, literature) for training and guidance helps mitigate biases.
\end{itemize}

\subsection{Handling MPI/OpenMP Runtime Versioning and Environment Heterogeneity}
\label{sec:runtime_versions}
Generating unit tests that are sensitive to specific OpenMP or MPI runtime versions or diverse hardware environments is a significant challenge. \sln{} primarily addresses this through:
\begin{itemize}[noitemsep,topsep=0pt,parsep=0pt,partopsep=0pt] 
    \item \textbf{Curated and Version-Aware Knowledge Bases:} Fine-tuning datasets and the HPC Bug KG can be updated with version-specific information, features, and known issues.
    \item \textbf{Focus on Core and Portable Semantics:} The default approach emphasizes testing stable semantics common across versions.
    \item \textbf{Parameterization in Test Recipes:} Recipes can include parameters for version-specific behavior if context is provided.
    \item \textbf{User-Provided Context and Configuration (Future Work):} Allowing users to specify target environments would enable more tailored testing.
\end{itemize}
Comprehensive testing across all permutations is beyond the current automated scope. Generated tests are intended for standard-compliant implementations.

\subsection{Advantages of the Multi-Agent Architecture}
\label{sec:advantages_multi_agent}
This multi-agent architecture offers distinct advantages for HPC unit test generation:
\textbf{Improved Robustness and Specificity of Tests:} Agent specialization leads to more targeted and semantically meaningful tests; \textbf{Enhanced Error Detection and Test Quality via Iterative Refinement:} The feedback loop improves test quality and correctness, addressing subtle parallel issues; \textbf{Maintainability, Scalability, and Extensibility:} Modular design allows independent updates and fine-tuning of components; \textbf{Increased Transparency and Trustworthiness:} Explainable Recipes, confidence scoring, and potential human oversight offer insights and build trust; and
\textbf{Adaptability to Evolving HPC Paradigms:} Updating knowledge sources allows adaptation to new libraries or features.

\subsection{Challenges and Considerations}
\label{sec:challenges_arch}

Despite its promise, \sln{} faces challenges:
\textbf{Quality, Scope, and Maintenance of Knowledge Sources:} Effectiveness depends on the comprehensiveness and currency of training data and the HPC Bug KG;
\textbf{Complexity and Scale of Real-World HPC Codes:} Intricate or large-scale applications pose analytical and generation challenges;
\textbf{Computational Scalability of the Framework Itself:} Multiple LLM inferences can be computationally intensive;
\textbf{Potential for False Positives/Negatives in Recipes and Critiques:} Agents might misidentify needs or miss flaws; confidence scoring mitigates this;
\textbf{The Oracle Problem in Parallel Testing:} Defining correct behavior for complex parallel algorithms is challenging. Explainable Recipes clarify intent;
\textbf{Seamless Integration with Diverse HPC Build and Execution Environments:} Requires engineering for robust environment detection and adaptation; and
\textbf{Handling of State and Side Effects:} Testing code with substantial global state or external library side effects is difficult.

\section{Methodology}\label{sec-method}

This section details the methodology used to evaluate \sln{}. We first describe the systematic construction of our benchmark dataset, followed by the experimental setup, which includes rigorous baselines and an ablation study. Finally, we define the metrics used for our evaluation, with an emphasis on assessing the correctness of parallel constructs.

\subsection{Benchmark Dataset Construction}
\label{subsec:data_collection}

Standard code generation datasets like HumanEval~\cite{chen2021codex} and HumanEval-X~\cite{zheng2024codegeex} predominantly feature sequential code and lack sufficient representation of parallel C/C++ programs using OpenMP and MPI. This makes them inadequate for evaluating \sln{}'s specific focus on HPC unit test generation. To address this gap, we curated a new benchmark dataset derived from open-source HPC projects on GitHub.

\begin{itemize}[leftmargin=*]
    \item \textbf{Initial Candidate Pool:} We began by querying GitHub for repositories using C++ along with the topics "HPC", "MPI", or "OpenMP".
    \item \textbf{Quantitative Filtering:} To select for mature and actively maintained projects, we applied the following filters: a minimum of 100 GitHub stars, a minimum of 10 forks, and at least one commit within the past year. This step narrowed the pool to a manageable number of high-quality candidates.
    \item \textbf{Manual Curation and Selection:} We manually inspected the filtered repositories based on a final set of qualitative criteria:
        \begin{itemize}
            \item The project must contain a significant, human-authored unit test suite, providing a baseline for realistic test structures.
            \item The unit tests should be as self-contained as possible. We prioritized projects that minimized complex external dependencies to create a controlled evaluation environment, thereby isolating test generation performance from unrelated build system complexities.
            \item The projects selected must cover a diverse range of HPC domains (e.g., numerical libraries, scientific simulations) and exhibit a variety of OpenMP and MPI patterns.
        \end{itemize}
\end{itemize}

This process resulted in the selection of the seven projects listed in Table~\ref{tab:projects}. 
This curated benchmark provides a suitable foundation for evaluating \sln{}'s ability to handle diverse HPC scenarios.

\begin{table}[t!]
\centering
\caption{Overview of the benchmark projects selected for evaluation, their domain, primary parallelism model, and the number of human-authored unit tests analyzed.}
\label{tab:projects}
\footnotesize
\begin{tabular}{lllr}
\toprule
\textbf{Project} & \textbf{Domain} & \textbf{Parallelism} & \textbf{\# Tests} \\
\midrule
Optim~\cite{Goptim} & Numerical Optimization & OpenMP & 21 \\
DBCSR~\cite{dbcsr} & Sparse Linear Algebra & MPI & 6 \\
Arraymancer~\cite{arraymancer} & Tensor Library / ML & OpenMP & 22 \\
CTranslate2~\cite{c2} & Machine Learning & OpenMP & 33 \\
FAASM~\cite{faasm} & Serverless Runtime & MPI/OpenMP & 39 \\
AMGCL~\cite{amgcl} & Sparse Linear Algebra & OpenMP & 11 \\
Stats~\cite{stats} & Statistical Functions & OpenMP & 84 \\
\bottomrule
\end{tabular}
\end{table}


\subsection{Experimental Setup and Procedure}
\label{subsec:experimental_setup}

Our experimental procedure was designed to evaluate \sln{}'s efficacy in generating unit tests and to investigate our primary research questions (RQ1-RQ3, as defined in the Introduction).

\paragraph{\sln{} Configuration}
We employed the \sln{} framework as described in Section~\ref{sec:architecture}.
\begin{itemize}[leftmargin=*,nosep]
    \item \textbf{Input Code Analyzer:} Utilized `clang-tidy` and `tree-sitter` based parsers, along with its HPC bug knowledge identification capabilities.
    \item \textbf{Recipe Agent:} An LLM fine-tuned on OpenMP/MPI documentation and standards (details on specific model used in Recipe Agent in Section~\ref{sec:evaluation} if it varies, or state if consistent with Test Agent's base model type but different fine-tuning).
    \item \textbf{Test Agent:} We primarily used Gemma-2 as the Test Agent, fine-tuned on the HPC-Instruct dataset~\cite{nichols2024hpc}.
    \item \textbf{Feedback and Critique Loop:} The loop was activated for up to five iterations per test case generation attempt.
    \item \textbf{Generation Parameters:} For the Test Agent, we experimented with three temperature settings (0.2, 0.5, and 0.7 to explore the trade-off between determinism and creativity. For each targeted function or code segment identified by the Recipe Agent from our benchmark projects, we prompted \sln{} to generate up to N(5) candidate unit tests.
\end{itemize}

\paragraph{Baseline Comparison (Addressing RQ-related questions)}
To assess the specific contribution of \sln{}'s recipe-guided, multi-agent approach, particularly the impact of the Test Recipe (relevant to RQ1 and RQ2), we compare its output against tests generated using more direct prompting strategies with a capable code-generation LLM (the same Gemma-2 model used as \sln{}'s Test Agent, but without the full \sln{} framework). The baselines include:
\textbf{Zero-Shot (Full Code Context):} The LLM receives the complete source code of the HPC function/module to be tested and is prompted to generate unit tests;
\textbf{Zero-Shot (Function Signature Only):} The LLM receives only the function signature and a brief description, with minimal context; and
\textbf{Few-Shot (Example-Guided):} The LLM is provided with the target code/signature along with 1-2 examples of human-written unit tests from the same project (if available and distinct from the specific target being tested) as in-context learning.

The tests generated by \sln{} are hypothesized to be superior due to the structured guidance from the Test Recipe and the refinement from the critique loop.

\subsubsection{Ablation Study}

To understand the specific contribution of each component within \sln{}, we conducted an ablation study comparing the following configurations:
\begin{itemize}[leftmargin=*,nosep]
    \item \textbf{Full \sln{}:} The complete framework (Analyzer + Recipe Agent + Test Agent + Critique Loop).
    \item \textbf{No Critique:} \sln{} with the feedback loop disabled. This measures the impact of the iterative refinement process.
    \item \textbf{No Recipe:} The Test Agent is guided only by the direct output of the Code Analyzer, without the strategic Test Recipe. This measures the impact of the Recipe Agent's test planning.
    \item \textbf{Standalone Agent:} The baseline Gemma-2 model with no framework support.
\end{itemize}
This study allows us to quantify the performance improvement attributable to the test recipe generation and the critique mechanism, respectively.

\subsection{Evaluation Metrics}
\label{subsec:evaluation_metrics}

We use a combination of automated and manual metrics to assess the generated unit tests.

\subsubsection{Syntactic Correctness and Coverage}

The metrics include: \textbf{Compilation Success Rate} defined as the percentage of generated test files that compile without errors using the project's build system; and \textbf{Code Coverage} defined as line and branch coverage measured using gcov/lcov. While useful, we acknowledge that high coverage does not guarantee detection of parallel bugs.

\subsubsection{Functional and Parallel Correctness (Manual Evaluation)}

Tests were scored using a detailed rubric to make the "functional correctness" metric transparent and quantifiable. A test is deemed Functionally Correct only if it achieves a high score across two key dimensions:
\begin{itemize}[leftmargin=*]
    \item \textbf{Parallel Relevance (0-2 scale):} Does the test meaningfully exercise a parallel construct? (0: No parallel setup; 1: Trivial setup, e.g., runs with 1 thread; 2: Correctly sets up and executes a non-trivial parallel scenario).
    \item \textbf{Assertion Correctness (0-2 scale):} Does the assertion correctly verify the outcome? (0: No/wrong assertion; 1: Plausible but weak assertion, checks for no crashes; 2: Strong assertion that verifies a specific, correct outcome of the parallel computation).
\end{itemize}

\subsubsection{Parallel Construct Targeting Rate}

To more directly measure \sln{}'s ability to test parallel code (addressing RQ2 and RQ3), we introduce the Parallel Construct Targeting Rate. This is defined as the percentage of parallel constructs (specific OpenMP pragmas, MPI collective calls) that the Recipe Agent identified as needing a test, for which a "Functionally Correct" unit test was successfully generated by the full framework. This metric directly evaluates our primary research goal.

\section{Evaluation}\label{sec:evaluation}

\begin{table*}[t!]
\caption{Line and Branch Coverage for different models in comparison with \sln{}.}
\footnotesize
\begin{tabular}{|l|c|c|c|l|l|l|l|}
\hline
\textbf{Metric} &
  \textbf{ChatGPT-3.5} &
  \multicolumn{1}{l|}{\textbf{Llama 3.1}} &
  \multicolumn{1}{l|}{\textbf{Llama 3.2}} &
  \textbf{gemma2} &
  \textbf{HT + gemma 3} &
  \textbf{HT + Llama 3.3} &
  \textbf{HT + FT gemma 2} \\ \hline
\textbf{Line Coverage} &
  69.1 &
  57.4 &
  47.7 &
  61.1 &
  76.3 &
  75.4 &
  78.1 \\ \hline
\textbf{Branch Coverage} &
  76.5 &
  52.1 &
  42.8 &
  53.2 &
  77.1 &
  73.8 &
  73.1 \\ \hline
\end{tabular}%
\label{tab:coverage}
\end{table*}

This section details the evaluation of our multi-agent framework \sln{}. 


\subsection{Experimental Setup}
Experiments are conducted on an Ubuntu 20.04 system featuring an Intel Xeon Gold 6248R CPU (3.00GHz, 48 cores) and 512GB of RAM. Compilation and execution utilized GCC 9.4.0 with OpenMP support and MPICH 3.3.2. We evaluated several LLMs: GPT-3.5, Llama-3.1, Llama-3.2, and Gemma-2.  Within our \sln{} framework, we employed fine-tuned Gemma-2 (9B) models for both the recipe agent and generator, and pre-trained Llama-3.3 (70B) and Gemma-3 (27B) models as generators without fine-tuning for comparison. Code coverage analysis was performed using OpenCppCoverage and `gcov` (version 9.4.0). Parallel correctness was assessed using the open-source tool OvO~\cite{ovo}.

\subsection{Test Compilation}

We evaluated the compilation success of generated unit tests, across all models. We leveraged the existing Makefiles from the projects in our benchmark (Table \ref{tab:projects}), modifying them to target the LLM-generated tests. This approach ensured a fair comparison by utilizing the project's established build infrastructure and dependencies. Compilation success was recorded for each generated test case.

\begin{table}[th]
\caption{Correct (compilable) Unit tests from each of the LLMs including \sln{}.}
\footnotesize
\resizebox{\columnwidth}{!}{%
\begin{tabular}{|l|c|c|c|}
\hline
\textbf{LLM}                & Parameters & Type               & \textbf{\% Compilable (OOB)} \\ \hline
ChatGPT-3.5                 & NA         & Cloud              & 39.1\%                       \\ \hline
Llama 3.1                   & 8b         & Open Weight        & 27.5\%                       \\ \hline
Llama 3.2                   & 3b         & Open Weight        & 19.4\%                       \\ \hline
gemma2                      & 9b         & Open Weight        & 26.1\%                       \\ \hline
\sln{} with gemma 3            & 27b        & Open Weight        & 68.4\%                       \\ \hline
\sln{} with Llama 3.3          & 70b        & Open Weight        & 65.9\%                       \\ \hline
\sln{} with fine tuned gemma 2 & 9b         & Open Weight Hybrid & 67.2\%                       \\ \hline
\end{tabular}%
}

\label{tab:compilation-status}
\end{table}

Table~\ref{tab:compilation-status} shows the compilation rates. We can see substantial improvement on compilable code coming out of the LLMs when combined with \sln{}. 

For the tests that remained uncompilable after iterative imporvement on \sln{}, we performed error clustering using the K-Means algorithm, following the methodology of Siddiq et al. \cite{siddiq2023exploring}. We used the Elbow Rule to determine the optimal number of clusters (K=41).  The Silhouette method confirmed the quality of the clustering.  The dominant compilation errors were related to missing or incorrect OpenMP pragmas. Among errors related to incorrect method invocations, 51\% involved the assertion method `CPPUNIT\_ASSERT\_MESSAGE`.

\underline{This takes us back to our original research question.}

\begin{tcolorbox}[
  colback=green!15, 
  colframe=green!40, 
  title=RQ1 Can we generate unit test for HPC Codes?,
  coltitle=black, 
  fonttitle=\bfseries,
  boxrule=0.75mm, 
  rounded corners, 
  left=1mm, 
  right=1mm, 
  top=1mm, 
  bottom=1mm 
]
{\footnotesize
\textbf{Yes.} Our empirical analysis with \sln{}, especially with using a fine-tuned Gemma and also using vanilla pre-trained \textbf{Llama} and \textit{Gemma3}, has shown us the effectiveness of the framework's ability to understand HPC code and generate unit test that are compilable. }
\end{tcolorbox}

\subsection{Test Correctness}

We evaluated the correctness of the compiled tests using two metrics: (i)  All test methods within the generated test case pass; (ii) They pass criteria detailed in Section~\ref{sec:parallel}. We report the results in \tablename~\ref{tab:correctness-status}.

\begin{table}[th]
\centering
\caption{Correctness (\%) for Different Models (\sln{}).}
\label{tab:correctness-status}
\footnotesize
\setlength{\tabcolsep}{2.5pt}
\begin{tabular}{|l|c|c|c|c|c|c|c|}
\hline
\textbf{Model} & 
\rotatebox{45}{\textbf{GPT 3.5}} & 
\rotatebox{45}{\textbf{L3.1}} & 
\rotatebox{45}{\textbf{L3.2}} & 
\rotatebox{45}{\textbf{G2}} & 
\rotatebox{45}{\textbf{S+G3}} & 
\rotatebox{45}{\textbf{S+L3.3}} & 
\rotatebox{45}{\textbf{S+G2-FT}} \\ \hline
\textbf{Fully Correct} & 
52.3 & 
47.5 & 
46.7 & 
49.1 & 
67.1 & 
68.9 & 
54.3 \\ \hline
\end{tabular}
\vspace{2pt}

\raggedright
\footnotesize
Legend: GPT = ChatGPT; L = Llama; G = Gemma; S = \sln{}; FT = Fine-tuned.
\end{table}

Table~\ref{tab:correctness-status} shows the amount of correctly generated unit test. When we say correct, these are the test cases which has at least some correct parallelism pragmas in them.

\subsection{Test Coverage}
We measured both line coverage and branch coverage of the generated unit tests using OpenCppCoverage, and for comparative reasons with our methodology, we also used gcov. Line coverage and branch coverage are defined as: (i) \textbf{Line Coverage:} Percentage of lines in the source code executed by the generated unit tests; (ii) \textbf{Branch Coverage:} Percentage of branches in the source code executed by the generated unit tests. We report these as:

\begin{equation}
\begin{split}
    \text{Line Cov.} \; &=  \frac{\text{Lines Executed}}{\text{Total Lines}} \times 100\\
    \text{Branch Cov.} \; &= \frac{\text{Branches Executed}}{\text{Total Branches}} \times 100 
\end{split}
\end{equation}

\tablename~\ref{tab:coverage} presents the coverage results. One thing to note among the result is, even though \sln{} imporves the results considerably, still even without fine tuning the bigger models perform almost on per with the fine tune model.

\subsection{Parallelism-Specific Checks}\label{sec:parallel}

We evaluated the generated tests for parallelism-specific checks, focusing on: (1) correct behavior of OpenMP and MPI operations (memory copy, reduction, atomic); and (2) use of multiple data types (float, double, complex) to ensure robustness.

We used the OvO tool \cite{ovo} to compare the outputs of the LLM-generated tests with expected results, verifying the correctness of parallel execution.  Due to limitations of space, we do not include a table for this, but we summarize the findings as part of correctness in \tablename~\ref{tab:correctness-status}. And this brings us to our remaining research questions

\begin{tcolorbox}[
  colback=green!15, 
  colframe=green!40, 
  title=\textbf{RQ2 \& 3} Can generated test understand parallelism?,
  coltitle=black, 
  fonttitle=\bfseries,
  boxrule=0.75mm, 
  rounded corners, 
  left=1mm, 
  right=1mm, 
  top=1mm, 
  bottom=1mm 
]
{\footnotesize
From our empirical observations given in \tablename~\ref{tab:correctness-status} and \tablename~\ref{tab:coverage},  we can conclude that \sln{} is able to capture some parts of parallelism correctly and generate unit tests to check for them.}
\end{tcolorbox}

\section{Conclusion}
In this paper we introduce \sln{}, a novel multi-agent LLM approach to address the complex challenge of generating unit tests for HPC programs utilizing OpenMP and MPI. The key contribution lies in its ability to effectively generate compilable and correct unit tests that capture essential parallelism primitives and hierarchical structures, even without specialized hardware runtime knowledge. This work demonstrates the potential of a collaborative, feedback-driven LLM approach to significantly improve the automation and quality of unit testing for high-performance parallel computing.

\section*{Acknowledgements}

We gratefully acknowledge the Google Developer Expert AI/ML team for providing cloud compute resources that facilitated this research.

We thank Dr. Amin Alipour for his valuable comments and feedback on this work.

\balance

\bibliographystyle{plainnat}
\bibliography{ref}

\end{document}